\documentclass[aps,prc,groupedaddress,showpacs,preprintnumbers,twocolumn]{revtex4}
\usepackage{
color}
\usepackage{epsfig}
\usepackage{longtable}
\begin{document}

\title{Mixing of the low-lying three- and five-quark $\Omega$ states with negative parity}

\author{C. S. An$^{1,2}$}\email{ancs@ihep.ac.cn}

\author{B. Ch. Metsch$^{2}$}\email{metsch@hiskp.uni-bonn.de}

\author{B. S. Zou$^{3,1}$}\email{zoubs@ihep.ac.cn}

\affiliation{1.Institute of High Energy
Physics and Theoretical Physics Center for Science Facilities,
Chinese Academy of Sciences, Beijing 100049, China\\
2.Helmholtz-Institut f\"{u}r Strahlen- und Kernphysik,
Universit\"{a}t Bonn, Bonn 53115, Germany\\
3.  State Key Laboratory of
Theoretical Physics, Institute of Theoretical Physics, Chinese
Academy of Sciences, Beijing 100190, China }

\thispagestyle{empty}

\date{\today}

\begin{abstract}

Mixing of the low-lying three- and
five-quark $\Omega$ states
with spin-parity quantum numbers $\frac{1}{2}^{-}$ and $\frac{3}{2}^{-}$
is investigated, employing an instanton-induced quark-antiquark pair
creation model, which precludes transitions between $s^{3}$ and $s^{4}\bar{s}$
configurations. Models with hyperfine interactions
between quarks of three different kinds, namely, one-gluon-exchange,
Goldstone-boson-exchange (GBE) and an instanton-induced interaction (INS) are
called OGE, GBE and INS models, respectively. Numerical results
show that the instanton-induced pair creation causes strong mixing between the
three- and five-quark configurations with spin $3/2$, and that this mixing
decreases the energy of the lowest spin $3/2$ states in all three different hyperfine
interaction models to $\sim 1750\pm50$~MeV. On the other hand, transition
couplings between $s^3$ and $s^{3}q\bar{q}$ states with spin $1/2$ caused by
instanton-induced $q\bar{q}$ creation is very small and the resulting mixing
of three- and five-quark configurations in the OGE and INS models is negligible,
while the mixing of the spin $1/2$ states in GBE model is not, but effects of
this mixing on energies of mixed states are also very
small. Accordingly, the lowest $\Omega$ states with negative parity
in all three hyperfine interactions models have spin $3/2$.

\end{abstract}


\maketitle

\section{Introduction}
\label{sec:intro}

Recently, we have studied the spectrum of low-lying $s^{3}Q\bar{Q}$ ~(where
$Q\bar{Q}=q\bar{q}\,,s\bar{s}$ for light and strange quark-antiquark pairs,
respectively) configurations with negative parity within an extended
constituent quark model with three different kinds of hyperfine interactions,
namely, one-gluon-exchange (OGE), Goldstone-boson-exchange (GBE) and
instanton-induced interactions (INS)~\cite{s^3}.  Experimental data about
$\Omega$ resonances is still very poor~\cite{pdg}: only four $\Omega$ states
were found~\cite{barns,Biagi:1985rn,Aston:1987bb,Aston:1988yn}, one being the
ground state $\Omega(1672)$, and all the other three states may also have
positive parity~\cite{Glozman:1995fu}.  A comparison of calculated results to
experimental data is therefore not very conclusive.  Compared to the
predictions of the masses of negative parity states in traditional three-quark
models, the lowest energy of $s^{3}Q\bar{Q}$ negative parity states is
expected to be $\sim180$ MeV lower~\cite{s^3}. This indicates that if we consider
$\Omega$ resonances as mixtures of three- and five-quark Fock components, then
the latter must be relevant for the properties of negative parity $\Omega$
resonances.

In the present paper, we shall study the mixing of $s^{3}$ and $s^{3}Q\bar{Q}$
configurations, which involves the investigation of transitions between three-
and five-quark Fock states.  For such transitions, the key ingredient is the
$Q\bar{Q}$ creation mechanism. Most widely accepted is the $^{3}P_{0}$
model~\cite{Le Yaouanc:1972ae}, which has been successfully applied to the
decays of mesons and baryons~\cite{Le Yaouanc:1973xz,Kokoski:1985is}, and this
was also employed to analyse the sea flavor content of the ground states of
the $SU(3)$ octet baryons~\cite{Santopinto:2010zza,An:2012kj}, as well as of
the spin and electromagnetic properties of
baryons~\cite{Bijker:2009up,Bijker:2012zza}.  In the $^{3}P_{0}$ model, the
$Q\bar{Q}$ pair is created with the quantum numbers of the QCD vacuum
$0^{++}$, which corresponds to $^{3}P_{0}$. There are also some other pair
creation models, for instance, string-breaking
models~\cite{Dosch:1986dp,Alcock:1983gb}, in which the lines of color flux
between quarks collapse into a string, the pair is created when the string
breaks and the created pair has as quantum numbers either
$^{3}P_{0}$~\cite{Dosch:1986dp} or $^{3}S_{1}$~\cite{Alcock:1983gb}. In
Ref.~\cite{JuliaDiaz:2006av}, the $Q\bar{Q}$ pair creation induced by a quark
confinement interaction was employed to investigate mixing between three- and
five-quark Fock components in the nucleon and the Roper resonance; in this
case the created $Q\bar{Q}$ also possesses the quantum numbers $^{3}P_{0}$.

In case of the low-lying $s^{3}Q\bar{Q}$ configurations with negative
parity, all the quarks and antiquarks are supposed to be relative $s$-waves,
and therefore the traditional $^{3}P_{0}$ pair creation mechanism can not
contribute. Accordingly we here employ an instanton-induced
interaction~\cite{'tHooft:1976fv,Shifman:1979uw,Petry:1985mn,Nachtsheim} for the pair
creation mechanism, since this interaction also can lead to the creation of
$Q\bar{Q}$ pairs with quantum numbers $^{3}S_{1}$ and $^{1}S_{0}$.  The
instanton-induced interaction has been used to describe the decays of
(pseudo)scaler mesons~\cite{Ritter:1996xh}.

The present paper is organised as follows. In Section \ref{sec:frame}, we
present our theoretical framework, which includes explicit forms of the
instanton-induced quark-antiquark pair creation mechanism.  Numerical results
for the spectrum of the states under study and the mixing of three- and
five-quark configurations in our model are shown in Section
\ref{sec:result}. Finally, Section \ref{sec:end} contains a brief conclusion.

\section{Theoretical Framework}
\label{sec:frame}

In the present model, to study mixing of the three- and five- quark
configurations, we describe the negative parity $\Omega$ states by
the Hamiltonian
\begin{equation}
  H=
  \pmatrix{ H_{3}  &  V_{\Omega_{3}\leftrightarrow\Omega_{5}}  \cr
    V_{\Omega_{3}\leftrightarrow\Omega_{5}}      &  H_{5}}
  \,,
  \label{ham}
\end{equation}
where $H_{3}$ is the Hamiltonian for a three-quark system and $H_{5}$ for a
five-quark system, and $V_{\Omega_{3}\leftrightarrow\Omega_{5}}$ denotes the
transition coupling between three- and five-quark systems. Note that in
principle the number of three and five quark configurations can exceed two.
Since the diagonal terms of~(\ref{ham}), the Hamiltonian $H_{3}$ for a
three-quark system has been discussed intensively in the literature and the
Hamiltonian $H_{5}$ for a five-quark system with the quantum numbers of
negative parity $\Omega$ resonances was recently developed in Ref.~\cite{s^3},
and will only briefly be reviewed here in Sec.~\ref{dia}. The non-diagonal
terms $V_{\Omega_{3}\leftrightarrow\Omega_{5}}$ will be explicitly discussed
in Sec~\ref{ndia}.

\subsection{Diagonal terms of the Hamiltonian}
\label{dia}

The Hamiltonian for a $N$-particle system in the non relativistic constituent quark model
is usually written as
 \begin{eqnarray}
   H_{N}&=&H_{o}+H_{\textit{\footnotesize hyp}}+\sum_{i=1}^{N}m_{i}\,,
\label{hn}
\end{eqnarray}
where $H_{o}$ and $H_{\textit{\footnotesize hyp}}$ represent the Hamiltonians
for the quark orbital motion and for the hyperfine interactions between
quarks, respectively, $m_{i}$ denotes the constituent mass of the $ith$
quark. The first term $H_{o}$ can be written as a sum of the kinetic energy
and the quark confinement potential as
\begin{equation}
  H_{o}=\sum_{i=1}^N {\vec{p}_i^2\over 2 m_{i}}+\sum_{i<j}^N
  V_{\textit{\footnotesize conf}}(r_{ij})\,.
  \label{ho}
\end{equation}
In~\cite{s^3} the quark confinement potential was taken to be
\begin{equation}
  V_{conf}(r_{ij})=-\frac{3}{8}\lambda_i^C\cdot\lambda_j^C
  \left[C^{(N)}(\vec{r}_i-\vec{r}_j)^2+V_0^{(N)}\right]\,,\label{conf}
\end{equation}
where $C^{(N)}$ and $V_{0}^{(N)}$ are constants. In principle these two
constants can differ for three- and five-quark configurations. The hyperfine
interactions between quarks $H_{hyp}$, as stated in~\cite{s^3}, can be
mediated by one gluon exchange, Goldstone boson exchange, or induced by the
instanton interaction. The forms of these three types of hyperfine
interactions in the three-quark system are explicitly given in the literature:
\cite{%
Glozman:1995fu,Isgur:1979be,Isgur:1978wd,Isgur:1977ef,%
Capstick:2000qj,Loring:2001kx,Blask:1990ez,Klempt:1995ku,Koll:2000ke}\,.
Those in the five-quark system with the quantum numbers of the $\Omega$ resonances
were explicitly discussed in~\cite{s^3} and will not be repeated here.

In the $N\leq2$ band \textit{e.g.} of the harmonic oscillator quark model there
are two $\Omega$ states with negative parity predicted by the three-quark
models corresponding to the first orbital excitation with $\ell = 1$\,: One
has spin $1/2$, the other
$3/2$~\cite{Glozman:1995fu,Chao:1980em,Pervin:2007wa}.  The energies are
obtained from the eigenvalues of Eq.~(\ref{hn}) in the case of $N=3$. The
results depend on the value of the strange constituent quark mass, the quark
confinement parameters $C^{(N=3)}$ and $V_{0}^{(N=3)}$ as well as the strength
of the hyperfine interaction.  To reduce free parameters, we just take the
values from~\cite{Chao:1980em} and~\cite{Glozman:1995fu} as matrix elements of
$H_{3}$ in the OGE and GBE models, respectively, \textit{i.e} in the OGE
model, $\langle H_{3}\rangle_{\frac{1}{2}^{-}}=\langle
H_{3}\rangle_{\frac{3}{2}^{-}}=2020$~MeV, and in the GBE model, $\langle
H_{3}\rangle_{\frac{1}{2}^{-}}=\langle
H_{3}\rangle_{\frac{3}{2}^{-}}=1991$~MeV.  In the INS model, since all three
quarks in $\Omega$ states are strange and thus the flavor state is symmetric,
the hyperfine interaction between quarks vanishes. Accordingly the matrix
elements of $H_{3}$ in this case only depend on the constituent mass of the
strange quark $m_s$ as well as $C^{(N=3)}$ and $V_{0}^{(N=3)}$\,.  If we adopt
the empirical values for $m_s$ and $C^{(N=3)}$ from~\cite{s^3}, and take
$V_{0}^{(N=3)}$ to be the tentative value which reproduces the mass of the
ground state $\Omega(1672)$, we find $\langle
H_{3}\rangle_{\frac{1}{2}^{-}}=\langle
H_{3}\rangle_{\frac{3}{2}^{-}}=1887$~MeV in the INS model.

Explicit matrix elements for $\frac{1}{2}^{-}$ and $\frac{3}{2}^{-}$ of the
sub-matrix $H_{5}$ in ~(\ref{ham}) were already listed in
Ref.~\cite{s^3}\,. In both cases $\langle
H_{5}\rangle_{\frac{1}{2}(\frac{3}{2})}$ are $4 \times 4$ matrices. Here we
just employ the results obtained within the OGE, INS and GBE models of
Ref.~\cite{s^3}.

\subsection{Non-diagonal terms of Hamiltonian}
\label{ndia}

The non-diagonal term $V_{\Omega_{3}\leftrightarrow\Omega_{5}}$ in the
Hamiltonian matrix~(\ref{ham}) describing the transition coupling between
three- and five-quark configurations depends on the explicit quark-antiquark
pair creation mechanism. The most commonly accepted mechanism for
quark-antiquark pair creation is the $^{3}P_{0}$ model~\cite{Le
Yaouanc:1972ae,Le Yaouanc:1973xz,Kokoski:1985is}.  In this model the created
quark-antiquark pair is in its first orbitally excited state, \textit{i.e.}
the $Q\bar{Q}$ pair has the quantum numbers $^{3}P_{0}$.  But in the present
case all quarks and antiquarks in the studied five-quark configurations are
assumed to be in their ground $s$-wave states and accordingly ${}^{3}P_{0}$
mechanism does not contribute to the coupling between $s^3$ and
$s^3\,Q\bar{Q}$ states considered here.

\begin{figure}[t]
\begin{center}
\includegraphics[scale=0.57]{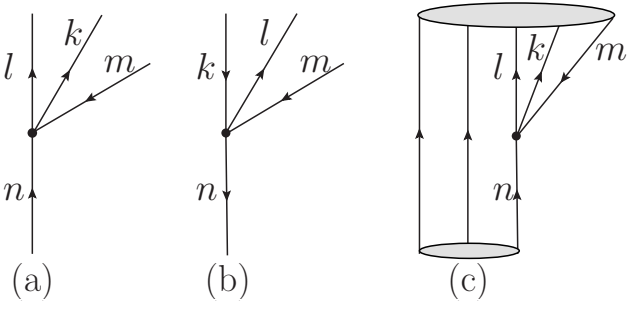}
\end{center}
\caption{\footnotesize (a).~Quark-antiquark pair creation from a quark;
(b).~Quark-antiquark pair creation from an antiquark; (c).~Transition
coupling of three- and five-quark configurations.
\label{fig}}
\end{figure}

Therefore we here adopt another quark-antiquark pair creation mechanism based
on a non-relativistic reduction of the amplitudes found from the
instanton-induced interaction.  This interaction was first proposed by 't
Hooft~\cite{'tHooft:1976fv} and developed by Shifman
et.~al.~\cite{Shifman:1979uw}, then Petry et.~al. applied it to the nuclear
structure~\cite{Petry:1985mn}.  Explicitly, the instanton-induced interaction
vertex for a quark-antiquark pair creation from a quark (antiquark) as shown
in Fig.~\ref{fig}~(a~(b)), can be written~\cite{Nachtsheim} in terms of normal
ordered products of creation and annihilation operators as
\begin{eqnarray}
 H_{q}&\colon=&-\frac{3}{16}g_{i}\epsilon_{ikl}\epsilon_{imn}\colon
 q_{l+}^{\dag}q_{k+}^{\dag}\left(\gamma_{0}\otimes\gamma_{0}+
 \gamma_{0}\gamma_{5}\otimes\gamma_{0}\gamma_{5}\right)\nonumber\\
&&\left(\mathcal{P}_{6}^{C}
+2\mathcal{P}_{\bar{3}}^{C}\right)q_{m+}q_{n-}\colon\,,
\label{ins}\\
H_{\bar{q}}&\colon=&-\frac{3}{16}g_{i}\epsilon_{ikl}\epsilon_{imn}\colon
q_{l+}^{\dag}q_{k-}^{\dag}\left(\gamma_{0}\otimes\gamma_{0}+
\gamma_{0}\gamma_{5}\otimes\gamma_{0}\gamma_{5}\right)\nonumber\\
&&\left(\mathcal{P}_{6}^{C}
+2\mathcal{P}_{\bar{3}}^{C}\right)q_{m-}q_{n-}\colon\,.
\label{insbar}
\end{eqnarray}
Here $H_{q}$ represents the pair creation from a quark, and $H_{\bar{q}}$ from
an antiquark, $g_{i}$ denotes strength of the instanton-induced interaction,
which has been discussed in~\cite{s^3}, $\mathcal{P}_{6}^{C}$ and
$\mathcal{P}_{\bar{3}}^{C}$ are projector operators on color $\bf{6}$ and
$\bar{\bf{3}}$ states, respectively, which are defined as
\begin{equation}
\mathcal{P}_{\bar{3}}^{C}=\frac{1}{2}\left(\mathcal{I}d-\Pi_{1,2}^{C}\right);
\mathcal{P}_{6}^{C}=\frac{1}{2}\left(\mathcal{I}d+\Pi_{1,2}^{C}\right)\,,
\label{pc}
\end{equation}
where $\mathcal{I}d$ denotes the identity, and $\Pi_{1,2}^{C}$ is the
permutation operator of two particles in color space.  Finally,
$\epsilon_{ikl}$ is the completely antisymmetric tensor acting on flavor
space. This precludes the creation of a quark-antiquark pair whose flavor is
the same as the quark $n$ or $l$. Consequently, in the present case, the
instanton-induced interaction does not mix $s^3$ and $s^{4}\bar{s}$
configurations. Note that the creation operator itself has only one free
parameter $g_{i}$, which should be the same as that for instanton-induced
hyperfine interactions between quarks~\cite{s^3}, therefore, no additional
parameter is introduced here.

It is obvious that $H_{q}$ in Eq.~(\ref{ins}) is the appropriate interaction
vertex we need for the present discussion in this paper. The normal ordering
in~(\ref{ins}) leads to two contributions: If $q_{m}$ has negative energy then
$q_{n}$ is an annihilation field operator and therefore
\begin{equation}
 \colon q^{\dag}_{l+}q^{\dag}_{k+}q_{m-}q_{n+}\colon=
 q^{\dag}_{l+}q^{\dag}_{k+}q_{m-}q_{n+}\,.
\end{equation}
If on the other hand $q_{m}$ is the annihilation field operator, then
\begin{equation}
 \colon q^{\dag}_{l+}q^{\dag}_{k+}q_{m+}q_{n-}\colon=
 -q^{\dag}_{l+}q^{\dag}_{k+}q_{n-}q_{m+}=q^{\dag}_{k+}q^{\dag}_{l+}q_{n-}q_{m+}\,.
\end{equation}
Thus the sign does not change if we simultaneously interchange
\begin{equation}
 k\longleftrightarrow l~~~~~ m \longleftrightarrow n\,.
\end{equation}
Therefore one obtains
\begin{eqnarray}
  H_{q}&=&-\frac{3}{8}g_{i}\epsilon_{ikl}\epsilon_{imn}
  q_{l+}^{\dag}q_{k+}^{\dag}\left(\gamma_{0}\otimes\gamma_{0}+
    \gamma_{0}\gamma_{5}\otimes\gamma_{0}\gamma_{5}\right)\nonumber\\
  &&\left(\mathcal{P}_{6}^{C}
    +2\mathcal{P}_{\bar{3}}^{C}\right)q_{m+}q_{n-}\,,
\label{ins'}
\end{eqnarray}
In addition, because of the total antisymmetry of the states, one can
eliminate the color projectors by replacing them by projectors in spin
space. After a non-relativistic reduction the quark-antiquark creation from a
quark by the instanton-induced interaction can be compactly written as
\begin{eqnarray}
  H_{q}&=&-\frac{3}{64}ig_{i}\epsilon_{ikl}\epsilon_{imn}\sum_{t=1}^{4}\sum_{\beta=0}^{3}
 h^{t}_{\beta}(k,l,m,n)\mathcal{D}_{t}(\xi_{k}^{\dag}\sigma_{\beta}\eta_{m})\nonumber\\
 &&(\xi_{l}^{\dag}\sigma_{\beta}\xi_{n})\,,
\label{rins}
\end{eqnarray}
where $\xi_{x}$ and $\eta_{x}$ represent quark and antiquark Pauli
spinors. The coefficients $h_{\beta}^{t}$ depend on the quark masses and are
given by
\begin{equation}\begin{array}{ll}
 h_{0}^{1}(k,l,m,n)=\frac{6}{m_{k}}-\frac{1}{m_{l}},
 & \hspace{0.5cm} h_{\beta>0}^{1}(k,l,m,n)=-\frac{1}{m_{l}},\\
h_{0}^{2}(k,l,m,n)=\frac{7}{m_{m}},
 & \hspace{0.5cm}h_{\beta>0}^{1}(k,l,m,n)=\frac{1}{m_{m}},\\
 h_{0}^{3}(k,l,m,n)=\frac{6}{m_{l}}-\frac{1}{m_{k}},
 &\hspace{0.5cm}h_{\beta>0}^{3}(k,l,m,n)=-\frac{1}{m_{k}},\\
 h_{0}^{4}(k,l,m,n)=-\frac{7}{m_{n}},
 &\hspace{0.5cm}h_{\beta>0}^{4}(k,l,m,n)=-\frac{1}{m_{n}},
 \end{array}
\end{equation}
where $\mathcal{D}_{t} = \sigma_{\alpha}\frac{\partial}{\partial x_{\alpha}}$
acting on the quark $t$ in~Eq.(\ref{rins}). For instance, for $t=3$,
$\mathcal{D}_{3}(\xi_{k}^{\dag}\sigma_{\beta}\eta_{m})
(\xi_{l}^{\dag}\sigma_{\beta}\xi_{n})=(\xi_{k}^{\dag}\sigma_{\beta}\eta_{m})
(\frac{\partial}{\partial
x_{\alpha}}\xi_{l}^{\dag}\sigma_{\alpha}\sigma_{\beta}\xi_{n})$.  Note that
here we defined $\sigma_{0}$ as the identity in spin space.  From
Eq.~(\ref{rins}), one finds that the created $q\bar{q}$ in the
instanton-induced pair creation model can have any of the quantum numbers
$^{3}P_{0}$, $^{1}P_{1}$, $^{3}P_{1}$, $^{1}S_{0}$ and $^{3}S_{1}$.  In the
case of $s^{3}\rightarrow s^{3}q\bar{q}$ states with negative parity
considered here only the latter two contribute.

The calculation of the transition $s^{3}\rightarrow s^{3}q\bar{q}$ then
involves the overlap between the residual three strange quarks in the
$s^{3}q\bar{q}$ configuration after $q\bar{q}$ annihilation and the initial
$s^{3}$ configuration, as shown in Fig.~\ref{fig}~(c).  Taking into account the
overlap factors and the overall symmetry, the transition coupling
$V_{\Omega_{3}\leftrightarrow\Omega_{5}}$ in (\ref{ham})
reads
\begin{eqnarray}
 V_{\Omega_{3}\leftrightarrow\Omega_{5}}&=&-\frac{9}{16}ig_{i}
 \epsilon_{ikl}\epsilon_{imn}\sum_{t=1}^{4}\sum_{\beta=0}^{3}
 \mathcal{C_{F}}\mathcal{C_{S}}\mathcal{C_{C}}\mathcal{C_{O}}\nonumber\\
 &&h^{t}_{\beta}(k,l,m,n)\mathcal{D}_{t}(\xi_{k}^{\dag}\sigma_{\beta}\eta_{m})
 (\xi_{l}^{\dag}\sigma_{\beta}\xi_{n})
\label{v35}
\end{eqnarray}
where $\mathcal{C_{F}}$, $\mathcal{C_{S}}$, $\mathcal{C_{C}}$ and
$\mathcal{C_{O}}$ are operators for the calculation of the corresponding
flavor, spin, color and orbital overlap factors, respectively.

\section{Numerical results}
\label{sec:result}

In the present treatment the matrix elements of transition coupling operator~(\ref{v35})
depend on two parameters, namely the instanton-induced interaction
strength $g^{\prime}$ and the oscillator parameter $\omega_{5}$
for the $s^{3}q\bar{q}$ configurations, if we adopt the constituent
mass of the strange quark and the oscillator parameter $\omega_{3}$
for the three-quark configuration as the empirical values from \cite{s^3}.
Notice that $\omega_{5}$ is from the orbital overlap factor.
In Sec.~\ref{con}, we present the numerical results obtained by taking
$g^{\prime}$ and $\omega_{5}$ as constant as previously used
in~\cite{s^3}. The dependence of the numerical results on the parameters is
the subject of Sec.~\ref{ncon}.

\subsection{Numerical results with fixed parameters}
\label{con}


\begin{table*}[ht]
\caption{\footnotesize Energies and the
corresponding probability amplitudes of three- and five-quark
configurations for the obtained $\Omega$ states in three
different kinds of hyperfine interaction models. The upper
and lower panels are for states with quantum numbers
$\frac{1}{2}^{-}$ and $\frac{3}{2}^{-}$, respectively,
and for each panel, the first row shows the energies in MeV, others show
the probability amplitudes.
\label{fix}}
\renewcommand
\tabcolsep{0.08cm}
\renewcommand{\arraystretch}{1.6}
\scriptsize
\vspace{0.5cm}
\begin{tabular}{c|ccccc|ccccc|ccccc}
\hline\hline
 &&&OGE&&& &&INS&&& && GBE &&\\
\hline
 $\frac{1}{2}^{-}$
&  2018       &   2149      &   2453      &   2656     &   2679
&  1796       &   1888      &   2030      &   2226     &   2432
&  1835       &   1892      &   1991      &   2018     &   2163
\\
\hline

$|3,\frac{1}{2}^{-}\rangle$
&  0.9961     &  -0.0774    &  -0.0394    &   0        &  -0.0167
&  0.1494     &   0.9854    &   0.0687    &   0.0425   &  -0.0096
&  0          &   0.2011    &   0.8964    &  -0.3951   &   0
\\

$|5,\frac{1}{2}^{-}\rangle_{1}$
&  0.0454     &   0.8428    &  -0.5332    &   0        &   0.0577
&  0.6650     &  -0.1563    &   0.7146    &   0.1097   &  -0.1031
&  1          &   0         &   0         &   0        &   0
\\

$|5,\frac{1}{2}^{-}\rangle_{2}$
&  0.0727     &   0.5326    &   0.8365    &   0        &  -0.1065
&  0.7318     &  -0.0592    &  -0.6630    &  -0.1066   &   0.1003
&  0          &   0.9676    &  -0.2446    &  -0.0623   &   0
\\

$|5,\frac{1}{2}^{-}\rangle_{3}$
&  0.0220     &   0.0068    &   0.1201    &   0        &   0.9925
&  0.0002     &   0.0301    &   0.1887    &  -0.9475   &   0.2563
&  0          &   0.1525    &   0.3697    &   0.9165   &   0
\\

$|5,\frac{1}{2}^{-}\rangle_{4}$
&  0          &   0         &   0         &   1        &   0
& -0.0036     &  -0.0089    &   0.0967    &   0.2775   &   0.9558
&  0          &   0         &   0         &   0        &   1
\\

\hline
 $\frac{3}{2}^{-}$
&  1727       &   2079      &   2366      &   2505     &   2519
&  1767       &   1991      &   2093      &   2193     &   2722
&  1773       &   1944      &   2010      &   2163     &   2166
     \\
\hline

$|3,\frac{3}{2}^{-}\rangle$
&  0.4989     &   0.8072    &  -0.3142    &   0.0299   &   0
&  0.8356     &  -0.0473    &   0.3243    &  -0.4353   &  -0.0692
& -0.6389     &   0.2538    &  -0.0594    &   0        &   0.7238
\\

$|5,\frac{3}{2}^{-}\rangle_{1}$
& -0.5556     &   0.2510    &  -0.3070    &  -0.7308   &   0
& -0.3013     &   0.7715    &   0.2032    &  -0.4772   &  -0.2120
&  0.6253     &   0.6984    &  -0.1904    &   0        &   0.2914
\\

$|5,\frac{3}{2}^{-}\rangle_{2}$
&  0.6651     &  -0.3885    &  -0.0029    &  -0.6377   &   0
&  0.2941     &   0.5539    &   0.1523    &   0.5306   &   0.5495
& -0.3637     &   0.6544    &   0.4155    &   0        &  -0.5165
\\

$|5,\frac{3}{2}^{-}\rangle_{3}$
&  0.0132     &  -0.3668    &  -0.8984    &   0.2414    &  0
& -0.3518     &  -0.3089    &   0.7586    &  -0.1450    &  0.4294
&  0.2617     &  -0.1395    &   0.8875    &   0         &  0.3528
\\

$|5,\frac{3}{2}^{-}\rangle_{4}$
&  0          &   0         &   0         &   0         &  1
& -0.0244     &  -0.0169    &  -0.5049    &  -0.5294    &  0.6812
&  0          &   0         &   0         &   1         &  0
\\

\hline
\hline
\end{tabular}
\end{table*}


In Refs.~\cite{s^3,Yuan:2012wz}, the empirical value for the strength of the
instanton-induced interactions between light and strange quarks was found to
be $g^{\prime}\simeq33.3$~MeV.  On the other hand, if we take the quark
confinement parameters equal, \textit{i.e.} $C^{(N=3)}=C^{(N=5)}$, we find a
relation between the oscillator parameters of three- and five-quark
configurations: $\omega_{5}=\sqrt{5/6}\omega_{3}$ and, correspondingly,
$\omega_{5}\simeq196$~MeV, as shown in~\cite{s^3}.  We denote the three-quark
configurations with quantum numbers $\frac{1}{2}^{-}$ and $\frac{3}{2}^{-}$ as
$|3,\frac{1}{2}^{-}\rangle$ and $|3,\frac{3}{2}^{-}\rangle$, respectively.
With the notation
\begin{eqnarray}
|5,\frac{1}{2}^{-}\rangle_{1}&=&|s^{3}q([4]_{X}[211]_{C}[31]_{FS}
[31]_{F}[22]_{S})\otimes \bar{q}\rangle\,,\nonumber\\
|5,\frac{1}{2}^{-}\rangle_{2}&=&|s^{3}q([4]_{X}[211]_{C}[31]_{FS}
[31]_{F}[31]_{S})\otimes \bar{q}\rangle\,,\nonumber\\
|5,\frac{1}{2}^{-}\rangle_{3}&=&|s^{3}q([4]_{X}[211]_{C}[31]_{FS}
[4]_{F}[31]_{S})\otimes \bar{q}\rangle\,,\nonumber\\
|5,\frac{1}{2}^{-}\rangle_{4}&=&|s^{4}([4]_{X}[211]_{C}[31]_{FS}
[4]_{F}[31]_{S})\otimes \bar{s}\rangle\,,
\label{num12}
\end{eqnarray}
for the five-quark configurations with spin $1/2$, and
\begin{eqnarray}
|5,\frac{3}{2}^{-}\rangle_{1}&=&|s^{3}q([4]_{X}[211]_{C}[31]_{FS}
[31]_{F}[31]_{S})\otimes \bar{q}\rangle\,,\nonumber\\
|5,\frac{3}{2}^{-}\rangle_{2}&=&|s^{3}q([4]_{X}[211]_{C}[31]_{FS}
[31]_{F}[4]_{S})\otimes \bar{q}\rangle\,,\nonumber\\
|5,\frac{3}{2}^{-}\rangle_{3}&=&|s^{3}q([4]_{X}[211]_{C}[31]_{FS}
[4]_{F}[31]_{S})\otimes \bar{q}\rangle\,,\nonumber\\
|5,\frac{3}{2}^{-}\rangle_{4}&=&|s^{4}([4]_{X}[211]_{C}[31]_{FS}
[4]_{F}[31]_{S})\otimes \bar{s}\rangle\,,
\label{num32}
\end{eqnarray}
for the five-quark configurations with spin $3/2$, the matrix elements of the
Hamiltonian~(\ref{ham}) in both cases are listed in Appendix~\ref{meh}.  Note
that the nonzero off-diagonal matrix elements in the sub-matrices $H_{5}$ are
caused by hyperfine interactions between quarks in the five-quark
configurations, as explicitly discussed in~\cite{s^3}. Diagonalization of
Eqs.~(\ref{engm1}) to~(\ref{engm6}) leads to the numerical results shown in
Table~\ref{fix}. In this table, we have ordered the states according to the
energy eigenvalue: The upper panel of the table shows energies of the states
with spin $1/2$, and the corresponding probability amplitudes of the three-
and five-quark configurations in these states, and the lower panel shows those
for the states with spin $3/2$.

From the upper panel of Table~\ref{fix} we conclude that in the OGE and INS
hyperfine interaction models the mixing between three- and five-quark
configurations with spin $1/2$ is very small, and even can be
negligible. Accordingly the resulting energies are very close to those
obtained in~\cite{s^3}, in which the effects of mixing between $s^{3}$ and
$s^{3}q\bar{q}$ were not included.  The mixing between three- and five-quark
$\Omega$ configurations obtained within the GBE hyperfine interaction model is
not so small that can be negligible: For instance, in the state with energy
$1991$ MeV, there is a $81\%$ three-quark component and $19\%$ five-quark
components. But also in this case the resulting energies are very close those
obtained in~\cite{s^3}.

In fact, absolute values of the transition matrix elements of $V_{35}$ in the
configurations with spin-parity $\frac{1}{2}^{-}$ are less than $20$ MeV, as
shown in Eqs.~(\ref{engm1}) to~(\ref{engm6}), which are tiny compared to the
diagonal matrix elements of the Hamiltonian matrix~(\ref{ham}).  These tiny
transition coupling matrix elements lead to tiny mixing between three- and
five-quark configurations with spin $1/2$ in the OGE and INS models.  However,
in the GBE model the situation is different: The mixing depends not only on
the couplings between the configurations, but also on the differences between
the diagonal matrix elements.  In the GBE model, as we can see in
Eq.~(\ref{engm5}), the diagonal matrix element of the third five-quark
configuration is close to that of the three-quark configuration, the
difference between the former and latter is only $19$~MeV. Therefore the
mixing of this five-quark configuration with the three-quark configuration is
not as small as that in the OGE and INS models. On the other hand, because the
diagonal energies of these configurations are close to each other, while
matrix elements of the non-diagonal transition coupling are small,
nevertheless the resulting energies are very close to those without mixing
between three- and five-quark configurations.

In case of the configurations with spin-parity $\frac{3}{2}^{-}$, as shown in
the lower panel of Table~\ref{fix}, mixing between three- and five-quark
$\Omega$ configurations in all the three hyperfine interaction models are very
strong.  Accordingly, the resulting energies differ substantially from those
without mixing between three- and five-quark configurations. The strongest
mixing is obtained within the GBE model, namely the state with energy $2166$
MeV: In this state, there is approximately $50\%$ three-quark component and
$50\%$ five-quark components. A very interesting result is that the lowest
states in all three models have energies lying in a narrow region around
$1750\pm25$ MeV.  This energy is significantly lower than the energies of the
lowest states with spin-parity $\frac{1}{2}^{-}$ in all three models.

Absolute values of the transition matrix elements of $V_{35}$ in the
configurations with spin-parity $\frac{3}{2}$ are in the range of $100\pm20$
MeV, which is much larger than the $20$~MeV couplings of configurations with
spin-parity $\frac{1}{2}^{-}$ and accordingly the mixing is much stronger than
in the $\frac{1}{2}^{-}$ case. In addition, the larger non-diagonal terms lead
to larger differences between the energies with and without mixing of three-
and five-quark configurations. As we have discussed in~\cite{s^3}, if we
ignore the mixing between three- and five-quark configurations, the lowest
state in the OGE model has spin $3/2$, but in the other two hyperfine
interaction models, the lowest states have spin $1/2$. If we take into account
the transition couplings between three- and five-quark configurations, the
lowest states in all three models have spin $3/2$.  Here, the lowest state in
the OGE model resulted partly by mixing between different five-quark
configurations caused by the OGE hyperfine interactions between quarks, and
partly by mixing between three- and five-quark configurations caused by the
instanton-induced $q\bar{q}$ pair creation. In the other two model the lowest
state is due to the action of the instanton-induced pair creation.

For both the $\frac{1}{2}^{-}$ and $\frac{3}{2}^{-}$ states, one may notice in
Table~\ref{fix} that there is at least one state that does not mix to other
states in the OGE and GBE models: These are the states
$|5,\frac{1}{2}^{-}\rangle_{4}$ and $|5,\frac{3}{2}^{-}\rangle_{4}$,
\textit{i.e.} the five-quark configurations with $s\bar{s}$.  As we discussed
in Sec.~\ref{ndia}, the instanton-induced pair creation interaction does not
lead to mixing between $s^{3}$ and $s^{4}\bar{s}$ configurations. Therefore,
in the OGE and GBE models, the state with the strange quark-antiquark pair
decouples from the others.  However, in the INS model, because of the
hyperfine interactions between quarks, there is no pure state with
$s\bar{s}$~\cite{s^3}, \textit{i.e.} the INS hyperfine interactions between
quarks leads to mixing between the $s^{4}\bar{s}$ and the $s^{3}q\bar{q}$
configurations. In the GBE model, there is another configuration that does not
mix with the others, namely the state $|5,\frac{1}{2}^{-}\rangle_{2}$. This is
because on the one hand the GBE hyperfine interactions between quarks cannot
mix this configuration with the other five-quark states, and on the other hand
the matrix element of the transition coupling $V_{35}$ between the three-quark
configuration and the $|5,\frac{1}{2}^{-}\rangle_{2}$ state vanishes, as shown
in Eq.~(\ref{engm5}).  In contrast, in the OGE and INS models, although the
transition coupling matrix elements between three- and five-quark
configurations vanish, the configuration $|5,\frac{1}{2}^{-}\rangle_{2}$ does
mix with other five-quark configurations, as we can see in Eqs.~(\ref{engm1})
and~(\ref{engm3}), so there is no pure state $|5,\frac{1}{2}^{-}\rangle_{2}$
in the OGE and INS hyperfine interaction models.

\subsection{Dependence of numerical results on parameters}
\label{ncon}

\begin{figure*}[t]
\begin{center}
\includegraphics[scale=0.9]{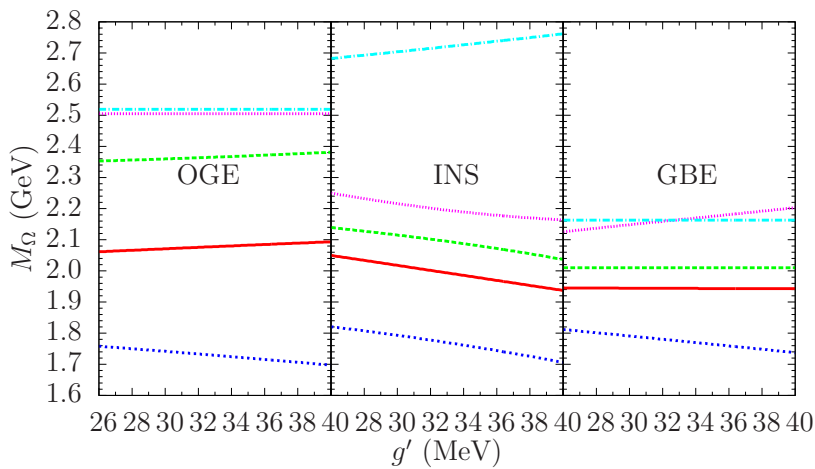}
\end{center}
\caption{\footnotesize (Color online) Energies of negative-parity $\Omega$ resonances with spin $3/2$ as function of
the instanton-induced interaction strength $g^{\prime}$. The figures from left to right
are the numerical results obtained within the OGE, INS and GBE models, respectively.
\label{gns}}
\end{figure*}

\begin{figure*}[t]
\begin{center}
\includegraphics[scale=0.9]{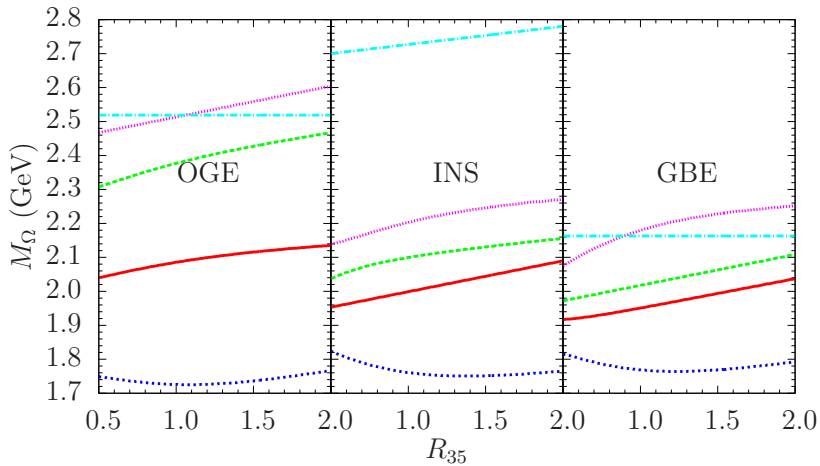}
\end{center}
\caption{\footnotesize (Color online) Energies of negative-parity $\Omega$ resonances with
  spin $3/2$ as function of $R_{35}$.
The figures from left to right
are the numerical results obtained within the OGE, INS and GBE models, respectively.
\label{r35}}
\end{figure*}

In Sec.~\ref{con}, we have shown the numerical results for a judicious choice
of parameters.  We still have to investigate whether the results are sensitive
to the interaction parameters and on the value of the oscillator parameter
$\omega_{5}$, which was just taken to be a tentative value in
Sec.~\ref{con}. Here we discuss the dependence of the energies on these
parameters.  Since the mixing between three- and five-quark configurations
with spin-parity quantum numbers $\frac{1}{2}^{-}$ was found to be very small,
we refrain from a discussion on the parameter dependence in this case.

Although the value for the instanton-induced interaction strength $g^{\prime}$
is an empirical one, it is the most important parameter determining the
transition between three- and five-quark configurations. Here we vary it by
$\pm20\%$ to demonstrate the dependence of the energies on its value in
Fig.~\ref{gns}.  The figures from left to right are the numerical results for
states with spin-parity $\frac{3}{2}^{-}$ obtained within the OGE, INS and the
GBE models.  As is evident from Fig.~\ref{gns}, the energies do not change
much in the OGE and GBE models within a range of $\pm40$ MeV.  But in the INS
model, the results show some more sensitivity to the coupling $g^{\prime}$,
mainly because the hyperfine interactions between quarks in the INS model also
depends on $g^{\prime}$.

An other important parameter in the matrix elements of the transition coupling
$V_{35}$ is the oscillator parameter $\omega_{5}$, which determines the
orbital overlap of three- and five-quark configurations. As discussed in
Sec.~\ref{con}, once we take $C^{(N=3)}=C^{(N=5)}$ in the quark confinement
potential, we get the ratio $R_{35}\equiv\omega_{5}/\omega_{3}=\sqrt{5/6}$,
but this is a tentative value only. In general, the color confinement strength
$C^{(N)}$ for three- and five-quark configuration could differ, so the value
of $R_{35}$ can also differ from $\sqrt{5/6}$. Now the oscillator parameter
$\omega_{N}$ reflects the size of the state studied If, for instance, we take
$R_{35}>1$, \textit{i.e.} $C^{(N=3)}< C^{(N=5)}$, this implies that the
five-quark configurations are more compact than the three-quark
configuration.  In this case, an intuitive picture for our model is going to
be like this: The three-quark state has a weaker potential; when quarks
expand, a $q\bar{q}$ pair is pulled out via the instanton-induced pair
creation mechanism and results in a $s^3q\bar{q}$ state with stronger
potential; the stronger potential leads to a more compact state which then
makes the $\bar{q}$ annihilate with a quark more readily leading to the $s^3$
state; this leads to constant transitions between these two states and
mixing. If, however, $R_{35}<1$, \textit{i.e.} $C^{(N=3)}>C^{(N=5)}$, the
picture is just opposite.  In
Refs.~\cite{An:2010wb,An:2011sb,Li:2005jn,Li:2006nm}, in order to reproduce
the electromagnetic and strong decays of nucleon resonances, both $R_{35}>1$
and $R_{35}<1$ have been suggested.
In~\cite{An:2010wb,An:2011sb,Li:2005jn,Li:2006nm} it was shown that the most
sensitive parameter is in fact the ratio
$2\omega_{3}\omega_{5}/(\omega_{3}^{2}+\omega_{5}^{2})$.  It is thus very
difficult to judge whether $R_{35}$ is less than $1$ or not. Accordingly, we
here vary the value of $R_{35}$ from $0.5$ to $2$ keeping the
instanton-induced interaction strength $g^{\prime}$ at the fixed empirical
value. The dependence of the energies of states with spin-parity
$\frac{3}{2}^{-}$ on $R_{35}$ are shown in Fig.~\ref{r35}. As in
Fig.~\ref{gns}, the figures from left to right are the numerical results
obtained within OGE, INS and GBE models.

One should notice that here we only want to show the dependence of the mixing
effects caused by transition couplings on the parameters, so we do not
consider the variation of the diagonal terms in Eqs.~(\ref{engm1})
to~(\ref{engm6}) with $\omega_{5}$. As we can see in Fig.~\ref{r35}, the
numerical results are somewhat sensitive to $R_{35}$ in all three interaction
models, the variation of the energies amount up to $160$~MeV. The energies of
the lowest states in the three models first decrease and then increase with
increasing values for $R_{35}$ and lie within the range $\sim1750\pm50$~MeV
The energies of the other states just increase with $R_{35}$, with the
exception of the blue-dash-dotted lines in the OGE and GBE models, which
represent the energy of the fourth five-quark configuration in these two
models, which does not mix with any others states in the OGE and GBE models,
so the energies of this state are independent of $R_{35}$.

\section{Conclusion}

\label{sec:end}

In this paper, we investigated the influence of the mixing of $s^{3}$ and
$s^{3}Q\bar{Q}$ $\Omega$ configurations on the energies of negative parity
$\Omega$ states with quantum numbers $\frac{1}{2}^{-}$ and $\frac{3}{2}^{-}$.
For the hyperfine interactions between quarks we investigated three
alternatives: the OGE, INS and GBE models.  The $Q\bar{Q}$ pair creation is
taken to be caused by the instanton-induced interaction. This mechanism has
the selection rule that it precludes $s\bar{s}$ creation from a strange quark
In another words, the instanton-induced interaction does not lead to mixing
between $s^{3}$ and $s^{4}\bar{s}$ configurations.

The matrix elements of the instanton-induced transition coupling in the spin
$1/2$ configurations are small, leading to negligible mixing between three-
and five-quark configurations with spin $1/2$ in the OGE and INS hyperfine
interaction models. In the GBE model, the spin $1/2$ state with strongest
mixing is composed of $\sim81\%$ three-quark and $\sim19\%$ five-quark
components mainly due an almost degeneracy of the corresponding unperturbed
three- and five-quark configurations.  Although the mixing of three- and
five-quark configurations with spin $1/2$ in the GBE model is not small, the
resulting energies are nevertheless very close those without mixing between
$s^{3}$ and $s^{3}q\bar{q}$, since the transition coupling matrix elements are
so small.

In the case of configurations with quantum numbers $\frac{3}{2}^{-}$, the
matrix elements of the instanton-induced transition coupling are much larger
and the resulting mixing between three- and five-quark configurations is very
strong. For instance, the spin $3/2$ state with the strongest mixing is
composed of $\sim50\%$ three-quark and $\sim50\%$ five-quark components.

The strong mixing between three- and five-quark configurations with spin $3/2$
decreases the energy of the lowest state appreciably in all the three hyperfine
interaction models: The lowest states with spin $3/2$ have an energy
$\sim1750\pm50$~MeV, which is lower than energies of all the spin $1/2$ states
obtained in the three different interaction models.
This is different from the results of previous models~\cite{Helminen:2000jb,Oh:2007cr}
without considering
the mixing between three- and five-quark configurations, which predicted the lowest
$\Omega$ excitation state to be of spin 1/2. To summarize: In all
interaction models the lowest states are found to be those with spin $3/2$ and
and lie at $\sim1750\pm50$~MeV. The lowest states differ in the three models:
Their major components are five-quark configurations ($\sim75\%$ and
$\sim64\%$, respectively) in the OGE and GBE models, whereas in the INS model,
the lowest state is mainly composed of the three-quark component
($\sim70\%$).

Very recently, BESII Collaboration at Beijing Electron Positron
Collider (BEPC) reported an interesting result that
$\psi(2S)\to\Omega\bar{\Omega}$ was observed with a branch fraction
of $(5\pm 2)\times 10^{-5}$~\cite{Ablikim:2012qp}. Now with the
upgraded BEPC, {\sl i.e.}, BEPCII, BESIII
Collaboration~\cite{Asner:2008nq} is going to take billions of
$\psi(2S)$ events, which is two orders of magnitude higher that what
BESII experiment got. If the lowest $\Omega$ resonance lies at
$\sim1750\pm50$~MeV, then it may be observed from from
$\psi(2S)\to\bar\Omega\Omega^*$ decays. Once an $\Omega^*$ resonance
is observed, its spin-parity quantum numbers can be obtained by a
partial wave analysis as demonstrated for the $N^*$ case
in~\cite{Ablikim:2004ug,Ablikim:2012zk}. Then the most interesting
result in the present paper that the lowest $\Omega$ resonance with
negative parity should have spin $3/2$ can be examined. However, it
seems to be difficult to distinguish the three different hyperfine
interaction models, since the predicted masses of the lowest state
in the three models are very close to each other, and the most
significant difference between the three models is that the
predicted probabilities of five-quark components are obviously
different, but it is not easy to be examined by the present
experimental measurements.

\section{Acknowledgements}
This work is supported by the National Natural Science Foundation of China under Grant
Nos. 11205164, 11035006, 11121092, 11261130311 (CRC110 by DFG and NSFC), the Chinese Academy of
Sciences under Project No. KJCX2-EW-N01 and the Ministry of Science and Technology of
China (2009CB825200).

\begin{appendix}

\section{Matrix elements of the Hamiltonian}
\label{meh}

The matrix elements of Hamiltonian~(\ref{ham}) in the configurations with
quantum numbers $\frac{1}{2}^{-}$ and $\frac{3}{2}^{-}$ read (numbers in units
of MeV)
\begin{widetext}
\begin{eqnarray}
\langle H^{OGE}\rangle_{1/2}=
\pmatrix{  2020.0   &     0       &  -18.5     & -13.1    &  0      \cr
           0        &     2235.0  &  -139.6    &  10.8    &  0      \cr
          -18.5     &    -149.6   &   2365.4   & -25.6    &  0      \cr
          -13.1     &     10.8    &  -25.6     &  2373.7  &  0      \cr
            0       &     0       &   0        &  0       &  2654.7 \cr} \,\label{engm1} \\
\langle H^{OGE}\rangle_{3/2}=
\pmatrix{  2020.0   &     115.8   &  -124.5    &  81.9    &  0      \cr
           115.8    &     2223.4  &   328.9    &  6.6     &  0      \cr
          -124.5    &     328.9   &   2095.0   & -68.0    &  0      \cr
           81.9     &     6.6     &  -68.0     &  2333.7  &  0      \cr
            0       &     0       &   0        &   0      &  2517.1 \cr} \,\label{engm2} \\
\langle H^{INS}\rangle_{1/2}=
\pmatrix{  1887.0   &     0       &  -18.5     &  -13.1   &  0      \cr
           0        &     1928.0  &  -121.5    &  -30.4   & -33.3   \cr
          -18.5     &    -121.5   &   1908.8   &   30.4   &  33.3   \cr
          -13.1     &    -30.4    &   30.4     &   2230.3 &  47.1   \cr
           0        &    -33.3    &   33.3     &   47.1   &  2411.0 \cr} \,\label{engm3}\\
\langle H^{INS}\rangle_{3/2}=
\pmatrix{  1887.0   &     115.8   &  -124.5    &   81.9   &  0      \cr
           115.8    &     2052.0  &  -113.3    &  -60.7   & -66.7   \cr
          -124.5    &    -113.3   &   2250.0   &   191.9  &  210.8  \cr
           81.9     &    -60.7    &   191.9    &   2159.0 & 188.5   \cr
           0        &    -66.7    &   210.8    &   188.5  & 2411.0  \cr}\,\label{engm4} \\
\langle H^{GBE}\rangle_{1/2}=
\pmatrix{  1991.0   &     0       &  -18.5     &  -13.1   &  0       \cr
           0        &     1833.6  &   0        &   0      &  0       \cr
          -18.5     &     0       &   1896.6   &  -16.2   &  0       \cr
          -13.1     &     0       &  -16.2     &   2010.0 &  0       \cr
            0       &     0       &   0        &   0      &  2161.6  \cr}\,\label{engm5}\\
\langle H^{GBE}\rangle_{3/2}=
\pmatrix{  1991.0   &     115.8   &  -124.5    &   81.9   &  0       \cr
           115.8    &     1896.6  &   0        &  -16.2   &  0       \cr
          -124.5    &     0       &   1990.2   &   0      &  0       \cr
           81.9     &    -16.2    &   0        &   2010.0 &  0       \cr
           0        &     0       &   0        &   0      & 2161.6   \cr} \,
\label{engm6}
\end{eqnarray}
\end{widetext}
\end{appendix}



\end{document}